# Bayesian estimate of position in mobile phone network

Aleksey Ogulenko[1], Itzhak Benenson[1*], Itzhak Omer[1], Barak Alon[2]

[1]Department of Geography and Human Environment,
Porter School of the Environmental and Earth Science, Tel Aviv University, Israel
[2]Partner Communications Company LTD
ogulenko.a.p@gmail.com, bennya@tauex.tau.ac.il, omer@tauex.tau.ac.il,
barak.alon1@partner.co.il, barakalon13@gmail.com

Abstract

The traditional approach to mobile phone positioning is based on the assumption that the geographical location of a cell tower recorded in a call details record (CDR) is a proxy for a device's location. A Voronoi tessellation is then constructed based on the entire network of cell towers and this tessellation is considered as a coordinate system, with the device located in a Voronoi polygon of a cell tower that is recorded in the CDR. If Voronoi-based positioning is correct, the uniqueness of the device trajectory is very high, and the device can be identified based on 3-4 of its recorded locations.

We propose and investigate a probabilistic approach to device positioning that is based on knowledge of each antennas' parameters and number of connections, as dependent on the distance to the antenna. The critical difference between the Voronoi-based and the real world layout is in the essential overlap of the antennas' service areas: the device that is located in a cell tower's polygon can be served by a more distant antenna that is chosen by the network system to balance the network load. This overlap is too significant to be ignored. Combining data on the distance distribution of the number of connections available for each antenna in the network, we succeed in resolving the overlap problem by applying Bayesian inference and construct a realistic distribution of the device location. Probabilistic device positioning demands a full revision of mobile phone data analysis, which we discuss with a focus on privacy risk estimates.

Keywords: Mobile phone positioning, Bayesian inference, Call Details Record, Location privacy

---

[*] Corresponding author

## 1. Individual's positioning based on the mobile phone data

Mobile phone data, as a source of information on the individual activities in time and space has a great potential for advancing all fields of the Geographic Information science and, specifically, for deeper understanding of population mobility and enhancing transportation and spatial planning (Berlingerio et al. 2013; Pinelli et al. 2016; Markovic et al. 2017). In high-resolution studies, these data are used for estimating and investigating individuals' daily mobility patterns (Gonzalez et al. 2008; Louail et al. 2014), travel patterns between the city core and periphery (Givoni, 2017), and mode-dependent commuting, such as of bikers and pedestrians (Xu et al. 2016; Kung et al. 2016; Bachir et al. 2019; Huang et al. 2019). When aggregated, mobile phone data assist general studies of urban and metropolitan dynamics (Calabrese et al. 2011; Razin and Charney, 2015), socioeconomic organization of cities (Cottineau and Vanhof, 2019) and urban land use planning (Pei et al. 2014), as well as monitoring of urban activities (Reades et al. 2009; Wu et al. 2020).

While different applications have different requirements in regard to the quality, resolution and level of data aggregation, the general tendency is to seek the highest possible resolution of the individual activities, in both space and time. High-resolution data are especially useful for transportation planning and management, like the analysis of public transport effectiveness, or establishing cycling lanes and pedestrian-only streets (Pinelli et al. 2016; Xu et al. 2016; Bachir et al. 2019). Critical in this regard is the ability to estimate the spatial location of mobile phone users. In the vast majority of studies, determining the location of a mobile device is based on the locations of the cell towers (base stations) of the mobile phone network. Namely, the units of the Voronoi coverage is constructed using the towers' coordinates and then the polygon of this coverage serve as basic units of the coordinate system: If, according the network's data record, the device d, at time moment t is connected to the antenna of the base station A, then the device is located within the Voronoi polygon $V_A$ of A. Typically, the exact position of d within the $V_A$ is not specified, while, sometimes the position of a base station itself is considered as a proxy of the position of d. Voronoi-based positioning of the mobile device is exploited in (Williams et al. 2015; Järv et al. 2017; Zufiria and Hernandez-Medina, 2018, 2019; Bonnetain et al. 2019; Cottineau and Vanhoof, 2019; Sotomayor-Gomez and Samaniego, 2020) and very few attempts are made to apply different approaches, like positioning the mobile device based on the frequency of the owners' visits (Wu et al. 2020). Some studies aim at improving the Voronoi-based positioning by applying a Kalman filter and incorporating the GPS data into the positioning algorithm (Hadachi and Lind, 2019) and, recently, considering the Voronoi polygons of towers' antennas instead of the polygons built for the towers (Bachir et al. 2019). Namely, a cell tower serves devices by means of three antennas, each covering a $120^o$ sector of the surrounding space and the Voronoi coverage is constructed based on the barycentres of the service areas of the antennas instead of the towers.

The Voronoi-based positioning is based, by definition, on the coverage of non-overlapping polygons. A solid criticism of this assumption was posed in several publications and conference presentations of Ricciato and co-authors (2017, 2020), who suggest that a solution for mobile device positioning must account for the overlap between the antennas' areas of service (Tennekes, 2018). However, this view still remains on the margins of the research attention. The goal of our study is to propose a framework and the software for mobile device positioning that accounts for the overlap between the service areas of the mobile phone network antennas.

The information about the overlap of the antennas' service areas is hidden in the distribution of the number of devices served by the antenna at different distances during a predefined period of time, which we call below the PRACH curve (see section 2). Based on this information, we propose a Bayesian estimate of the device position that does account for antennas' overlap. As we demonstrate, the traditional Voronoi-based positioning results in significantly biased and unrealistically precise estimates, and the latter has important consequences for the privacy-related aspects of the mobile phone data.

Sections 2 of the paper presents the details of a mobile phone network that are essential for our study. Section 3 of the paper presents Bayesian estimates of the mobile phone position. Section 4 presents a comparison between the Voronoi-based and Bayesian estimates. In Section 5, we consider the consequences of the proposed positioning methodology, which brings significant additional uncertainty to our knowledge about mobile device positioning.

The software developed in this project is available for free download at the https://github.com/grauwelf/mob-bayes-clouds. It's important to note that our approach is based on the standard aggregate data collected by every mobile network operator for the purposes of the network maintenance.

## 2. Mobile phone network data description

The most abstract representation of a mobile phone network (MPN) serving individual devices of a third and fourth generation is as follows: an MPN consists of two sets of antennas, indoor and outdoor. Outdoor antennas have a constant orientation and sector of service, and are located on cell towers, each bearing several outdoor antennas. The capacity of the outdoor antenna, that is the number of devices that it can simultaneously serve, is constant; while different antennas may have different capacities. Outdoor antennas are capable of serving mobile phones of all generations. Typically, there are 3 antennas on a tower, each covering a $120^o$ sector. Technical characteristics of outdoor antennas makes it possible to serve mobile devices at a distance of up to ~30 km, and Figure 1 presents examples of antenna service areas. Antennas sectors of service are divided into rings - Trip-Time Bands (TTBs) that are explained later in the paper. Indoor antennas are located inside buildings and aim at serving devices at a distance of hundreds of meters at most. Their capacity is much lower than that of outdoor antennas.

The major goal of the MPN is to supply maximum possible quality of service to customers' devices. The number and location of cell towers is thus an outcome of the compromise between complicated technical and legal limitations. A real-world MPN is never steady and its configuration constantly changes. At the macro-scale, the network is always undergoing maintenance and repair. On the micro-scale, the network is constantly adapting to the state of the propagation medium, fluctuation of the demand and load, interference between antennas of the same or another MPN, and so on. The related algorithms of the telecommunication system are inherently non-deterministic and, for example, it is impossible to predict what frequency band will be chosen for connection between an antenna and mobile phone.

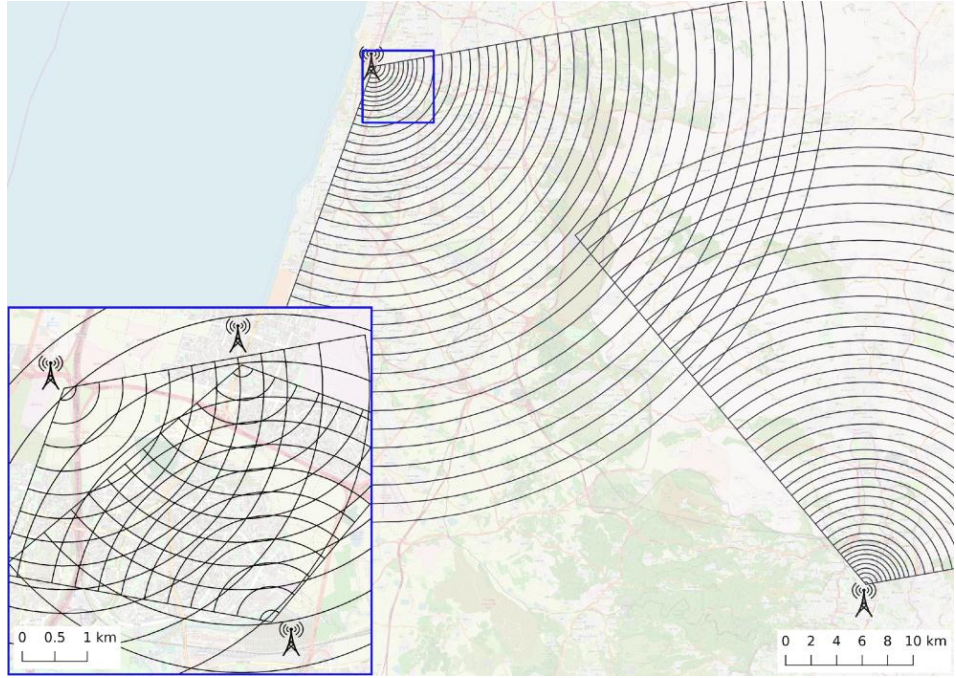

Figure 1: Main map shows overlapping sectors of service for two antennas located a large distance apart. Inset: The view of the overlap of sectors of service for three antennas located close by.

The maximum possible quality of service to the customers' devices is achieved by instantaneous balancing between the devices' requests and antennas' load. Upon a device request, the MPN software considers several antennas around, including nearby indoor antennas, as candidates for serving the request. The antenna that is chosen by the MPN for service is often not the closest one. Moreover, different antennas can service requests from a stationary device, and the antenna can even be switched during the same phone call.

Antennas' service areas with their TTBs are presented in Figure 1. To balance the antennas' loads, the MPN is able to estimate the distance between the antennas and a device. The distance is estimated based on the signal round-trip-time (RTT) or signal strength, and may be recorded in the device's call detail record (CDR). The estimate of the distance is imprecise and is considered by the trip-time bands (TTB) of the antenna's service sector. As shown in Figure 1, the width of the TTB increases with the growth of the distance from the antenna. The width of TTB rings is not a round number. The width of the ring closest to an antenna is about 200 m, the next closest ones are ~ 400 m, after which the width increases to ~ 700 m, and the most distant are ~1300 m. TTBs for each MPN antenna are known, do not change in time, and the increase in their width with the distance is similar for the majority of them.

There are several types of connection sessions (voice calls, SMS, WiFi, sighting). Let us consider the simplest example of a voice call. The CDRs of the voice call contain the time of connection, ID of the antenna, and cell tower ID. Usually, the connection start and end are recorded, and, typically, there are more than two CDRs recorded during the talk. If the call was managed by several antennas, the antenna ID and cell tower IDs of all antennas involved, plus the moments of re-connection are recorded. Ideally, the spatial components of the CDR contain, besides the antenna and cell tower IDs, the sequential number of the distance ring (Table 1).

| Table 1: Schematic representation of the CDRs of a short voice call ||||||
|---|---|---|---|---|---|
| **Device ID** | **Start timestamp** | **End timestamp** | **Tower ID** | **Antenna ID** | … |
| …D86BA7 | 2020-01-22T17:41:42.000 | 2020-01-22T17:43:30.000 | …7DC5 | …E002 | |
| …D86BA7 | 2020-01-22T17:43:30.000 | 2020-01-22T17:43:52.000 | …4EBD | …D26B | |
| …D86BA7 | 2020-01-22T17:43:52.000 | 2020-01-22T17:43:57.000 | …7DC5 | …E002 | |
| …D86BA7 | 2020-01-22T17:43:57.000 | 2020-01-22T17:44:54.000 | …7DC5 | …E002 | |
| …D86BA7 | 2020-01-22T17:44:54.000 | 2020-01-22T17:44:55.000 | …BC1A | …640B | |
| …D86BA7 | 2020-01-22T17:44:55.000 | 2020-01-22T17:49:09.000 | …BC1A | …8215 | |
| …D86BA7 | 2020-01-22T17:49:09.000 | 2020-01-22T17:49:46.000 | …BC1A | …8215 | |

Usually, the CDR data that are available to researchers are somehow aggregated or censored and, typically, the CDR data contain the cell tower IDs, but not the antenna IDs, and time of the connection is somehow rounded. The major disadvantage of the aggregate/censored data is the lack of the information on the distance to the device. The latter seems to be the major reason for the broad view that Voronoi tessellation, based on the locations of the MPN cell towers, may serve as a proxy for the device location. Namely, it is assumed that at the time moment recorded in the CDR, the device is located within the tessellation polygon of the tower that is recorded in this CDR (Candia et al. 2008; Song et al. 2010; Csáji et al. 2012; De Montjoye et al. 2013; Bonnel et al. 2015; Kalatian and Shafahi, 2016). This assumption has far-reaching consequences, especially in regards to user location privacy, and if this is true, then only 3 – 4 locations of a device are sufficient to identify the specific device with close to a 99% probability (De Montjoye et al. 2013).

Voronoi-based positioning is extensively used and well-studied. It is unambiguous, computationally effective, and is clearly intuitive. It accepts as a self-evident fact that the tower placed in the polygon's centroid absolutely dominates over the whole area inside the polygon. However, Voronoi tessellation does not account for the overlap of service areas of multiple antennas caused by the basic physical aspects of the MPN structure and workflow (Zhang, 2017):

- *Cross-slot interference*: Communication between a user device and base station consist of repeated sequences of periods: *downlink* period (base station → user device), *silent guard* period, and *uplink* period (user device → base station). For the given climate and environment, the downlink signal from a distant base station/antenna may arrive with very low propagation loss yet with a significant delay, hitting into the uplink period of the target base station. Typical scenarios include base stations on top of hills around large city, base stations in different cities, or those separated by a large water body. While normal communications are performed over distances up to 30 km, cross-slot interference can result in communication with the antennas at distances up to 200-300 km.
- *Uplink interference*: Uplink from a user device at a location that can be served by several towers, such as on the border between cellular cells, will cause interference to adjacent towers. Devices with bad radio-frequency conditions that unsuccessfully try to get access to the chosen station and transmit high power signals can cause high noise conditions. The latter leads to access failures and results in heterogeneity of communication quality inside the coverage area.
- *Doppler shifts*: A fast-moving user device causes Doppler shifts (offset of radio-frequency) in the uplink signal received by a base station. The user device then synchronizes to a shifted downlink signal, and its next uplink will be shifted more, and so on. In this way, Doppler shifts cause fundamental performance degradation. Typical scenarios refer to high-speed trains or highways with base stations installed along the road. To manage Doppler shifts, the MPN has to be tuned in order to compensate for the high-frequency offsets, and this also leads to the heterogeneity of communication quality inside the coverage area.

Recent doubts about the adequacy of the tower-based Voronoi partition resulted in essential modifications. Two important examples are *sectors Voronoi* partition (Bachir et al. 2019) that is constructed based on the barycentres of the centroids of cells within the antenna's sector and, especially, *section tessellation* technique (Ricciato et al. 2017) that uses service coverage maps to account for the overlap of the cell towers' coverage areas and shows significant gain of spatial accuracy in simulated scenarios with the synthetic population and MPN coverage.

## 3. Bayesian inference of the mobile phone position

### 3.1. Bayesian estimate of antenna's service area

To account for probabilistic nature of the MPN service, we apply the Bayesian approach to the device's location. To establish the model, we assume that the following components are fully defined:

1) Network layout — location of every cell tower, location of antennas on the towers; azimuth and TTBs of antennas.

2) A posteriori distribution, for each antenna, of the number of connections by antenna's TTBs – PRACH curves, constructed over a period of time that is long enough to obtain stable estimates. In what follows, we consider PRACH curves constructed for one month.

Let us consider a device D located inside the coverage area of a set of antennas $A = \{A_1, A_2, ..., A_n\}$ and estimate the probability that D is located at a given point $\bar{p}(x, y)$, given its current connection is carried by the antenna $A_j \in A$. Let the TTBs of the $A_j$, in order of their distance from the antenna, be $B(A_j) = \{b_{j_1}, b_{j_2}, ..., b_{j_l}\}$, where *l* is a total number of TTBs for the $A_j$. Each antenna is characterized by the monthly number of connections with the devices located within each TTB. This statistic is called below a *PRACH curve*, from the Physical Random Access Channel procedure used by a device to initiate contact with a base station (Korhonen, 2003, p. 340).

A PRACH curve is defined by two factors — spatial distribution of the population's communication activity and the overlap between antennas' areas of service. In this study we use the PRACH curves supplied to the mobile phone operator by a third-party company. According to the industry standards, the curves are estimated up to the distance of 32 km, and more distant connections are not included. Figure 2 presents PRACH curves for three antennas that represent typical but very different kinds of this curve.

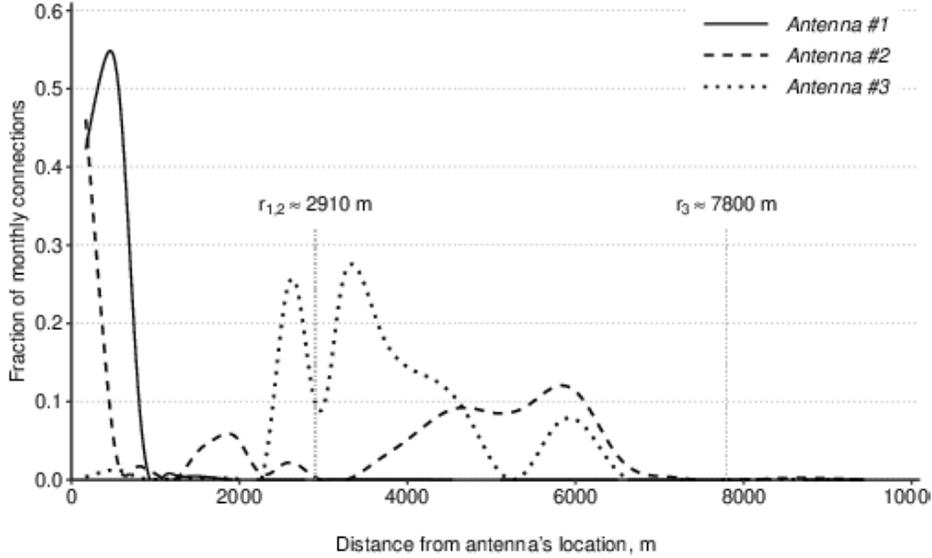

Figure 2: Monthly PRACH curves for three antennas. $r_{1,2}$ and $r_3$ denote circumcircle's radii of the Voronoi polygon (see below) for the towers of these antennas.

As can be seen, the PRACH curves in Figure 2 are very different. Antenna #1 has one peak close to its vicinity, antenna #2 has an essential number of connections at the distance interval of 4–6 km, and antenna #3 serves many devices within the 2–7 km distance interval.

We consider device position at the resolution of the discrete square grid $\Delta = \{\Delta_1, \Delta_2, \ldots, \Delta_s\}$ and estimate the probability that the device served by the antenna $A_j$ is located in a grid unit $\Delta_i$ applying Bayes theorem:

$$P(\Delta_i|A_j) = \frac{P(A_j|\Delta_i)P(\Delta_i)}{\sum_{k=1}^{s} P(A_j|\Delta_k)P(\Delta_k)}, i = \overline{1,s}, j = \overline{1,n}. \qquad (1)$$

Here $P(A_j|\Delta_i)$ is the probability that the device is served by antenna $A_j$ given the device is located in the grid element $\Delta_i \in \Delta$ and $P(\Delta_i)$ is an *a priori* number of devices in $\Delta_i$.

We estimate $P(A_j|\Delta_i)$ based on overlap between the grid elements $\Delta_i$ and PRACH curve of all antennas $A_j$ which TTBs overlap $\Delta_i$. Given a time moment t, a device that is located in $\Delta_i$ can be connected at t to one antenna $A_j$ only. Let us consider the position of $\Delta_i$ in respect to $B(A_j)$ and let $b_{j_k}$ be one of the $A_j's$ TTBs for which $b_{j_k} \cap \Delta_i$. In what follows we assume that the probability that the connection was established from $\Delta_i$ to $A_j$ is proportional 1) to the area of intersection $b_{j_k} \cap \Delta_i$ and 2) to the fraction of connections carried by $b_{j_k}$ among all connections established from $\Delta_i$ to all antennas which TTBs cover $\Delta_i$. The latter is estimated using *a posteriori* distribution of network activity.

Decomposing $P(A_j|\Delta_i)$ in a sum over all TTBs from $B(A_j)$ we obtain:

$$P(A_j|\Delta_i) = \sum_{b_{j_k} \in B(A_j)} P(A_j|\Delta_i) =$$
$$= \sum_{b_{j_k} \in B(A_j)} \frac{area(b_{j_k} \cap \Delta_i)}{area(\Delta_i)} \times \frac{\# \, of \, connections \, carried \, by \, b_{j_k} \, in \, \Delta_i}{Total \, \# \, of \, connections \, in \, \Delta_i} \quad (2)$$

An *a priori* distribution of devices' locations $P(\Delta_i)$ can be considered, in respect to the prior information, in two ways (Williamson, 2010): From the "objective" point of view, we should avoid any prior assumptions about $P(\Delta_i)$ and thus assume the prior distribution is uniform, $P(\Delta_i) = constant$. That is, we can reduce the terms $P(\Delta_i)$ and $P(\Delta_k)$ in (1). "Subjective" estimation assumes that we have some independent knowledge about prior location distribution. For example, we can assume that priors are proportional to the population of a grid cells and the population information is available from the census. In this case, we would ignore information on the devices that are located in the grid cells temporarily, and exclude unpopulated grid cells from the further calculations. On this basis, we prefer an "objective" view.

### 3.2. Examples of Bayesian estimates of antenna's service area

Applying Bayesian estimates (1) – (2), we obtain a set of grid cells – a "cloud" of possible location of a device that is registered at a given antenna. This study is based on the information on 22007 antennas of Partner Communications Company LTD ("Partner") MPN that serves the entire area of Israel. Each antenna of this MPN is technically able to serve devices up to a distance of 32 km, and its PRACH curve is presented by 40 TTBs. For each antenna we possess knowledge on its location (via the cell tower location), azimuth and monthly PRACH curves, and, based on that, have estimated *a posteriori* distributions of each antenna's connections, based on an "objective" view of the priors.

MPN's coverage area was discretized into a grid of 250 × 250 m cells, and the total number of these cells for Israel is 360139. To reduce computational cost, we consider only those TTBs whose average monthly density of connections is at least 10 per 250 × 250 m grid cell per month, that is, 160 connections per 1 km$^2$ per month. This limitation resulted in excluding 0.02% of all connections and the remaining 99.8% of the overall 42×10$^9$ connections, estimated as a sum of connections over all PRACH curves, were used for the Bayesian inference. According to this criterion, 335195 (93.1%) of the grid cells are further included into at least one probabilistic cloud. In what follows, we consider the outcomes that are based on these 99.8% of the observations as 100%. All calculations were performed in the *PostgreSQL* database with the use of the *PostGIS* GIS extension for performing spatial operations.

Figure 3 shows probabilistic "clouds" for eight antennas. To present the cloud for antenna $A_j$ for a given cumulative probability p (*p-cloud* of the antenna), we sort all grid cells for which probability $P(\Delta_i|A_j)$, estimated in (1) is positive, and consider a minimal set of cells with the highest $P(\Delta_i|A_j)$ that comprise a total share greater than or equal to p. Figure 3 illustrates the important fact that antennas essentially differ in regards to the certainty of device positioning.

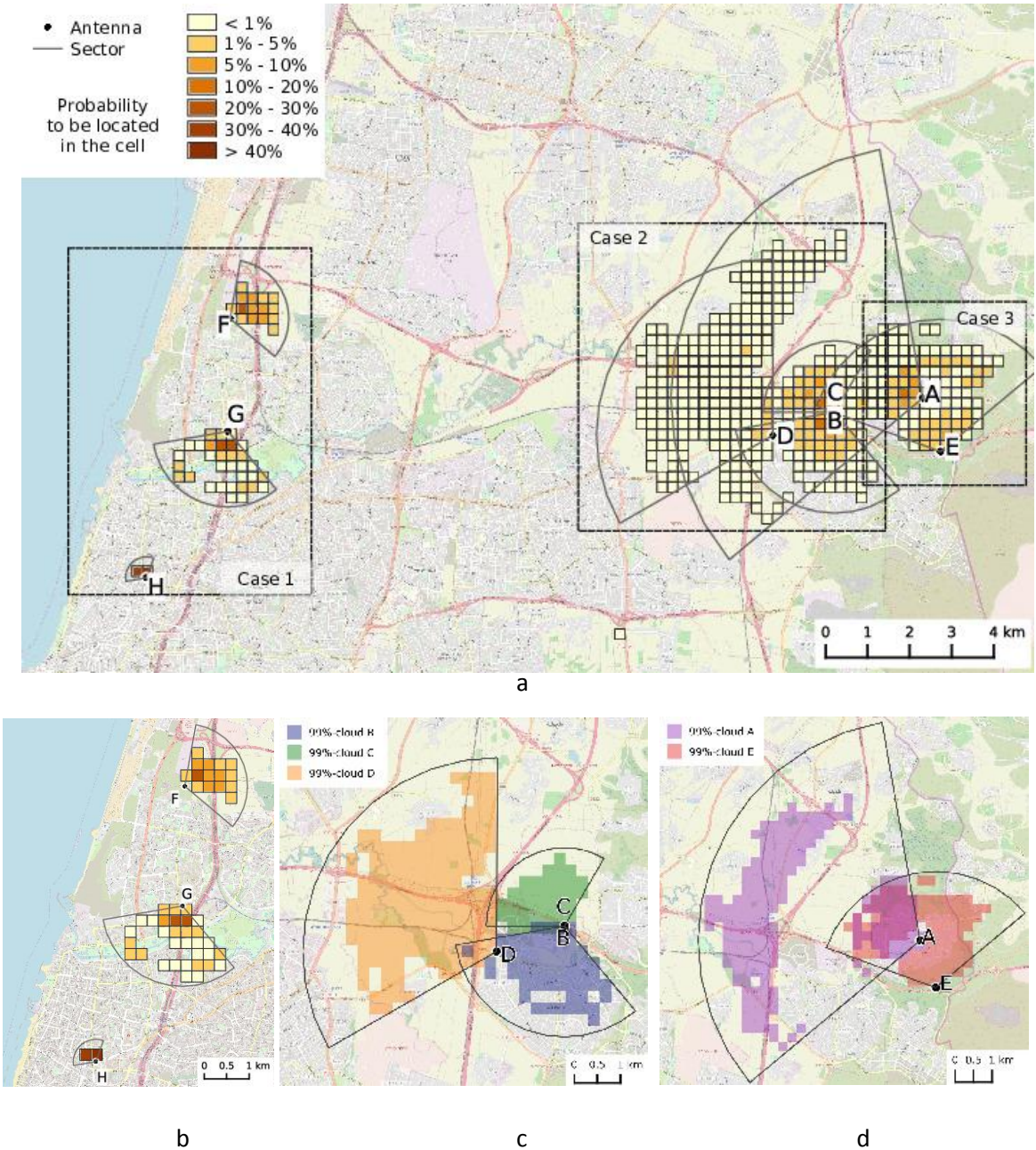

Figure 3: 99% Location clouds for eight antennas A – G, that are positioned on different cell towers in the Tel Aviv Metropolitan Area, at a resolution of 250x250 m grid. (a) General view; (b) Very small cloud H, cloud of an average size F, discontinuous cloud G; (c) Three clouds of antennas located at adjacent towers with very small overlap; (d) Cloud of antenna E covers the core part of the discontinuous cloud of antenna A.

Let us now extend the Bayesian approach towards the location cloud of the cell tower and compare the probabilistic location to the Voronoi-based one.

### 3.3. Bayesian estimate of the device location based on the cell tower data

We define a probabilistic location cloud for a tower as a union of the probabilistic location clouds of the tower's antennas. Let us consider a cell tower $M$ that is equipped with antennas $A_{j_1}, A_{j_2}, \ldots, A_{j_p} \in A$. Applying Bayes theorem in the same way as in (1) – (2), we obtain

$$P(\Delta_i|M) = P\left(\Delta_i | A_{j_1} \cup A_{j_2} \cup \ldots \cup A_{j_p}\right) =$$

$$= \frac{P(A_{j_1} \cup A_{j_2} \cup \ldots \cup A_{j_p}|\Delta_i)P(\Delta_i)}{P(A_{j_1} \cup A_{j_2} \cup \ldots \cup A_{j_p})} = \frac{\sum_{k=1}^{p} P(A_{j_k}|\Delta_i)P(\Delta_i)}{\sum_{k=1}^{p}\sum_{l=1}^{s} P(A_{j_k}|\Delta_l)P(\Delta_l)}, \quad i = \overline{1, s} \quad (3)$$

The $P(A_{j_1} \cup A_{j_2} \cup \ldots \cup A_{j_p}|\Delta_i)$ denotes a probability that the device D, located at the grid element $\Delta_i \in \Delta$, will be served by some antenna of the cell tower $M$. Since, at a given time moment, the device can be connected to one antenna only, we can consider this probability as a sum of a posteriori probabilities $P(A_{j_k}|\Delta_i)$ over all the tower's antennas. In this way, we construct an estimation of the PRACH curve for the tower as combination of the PRACH curves for the tower's antennas.

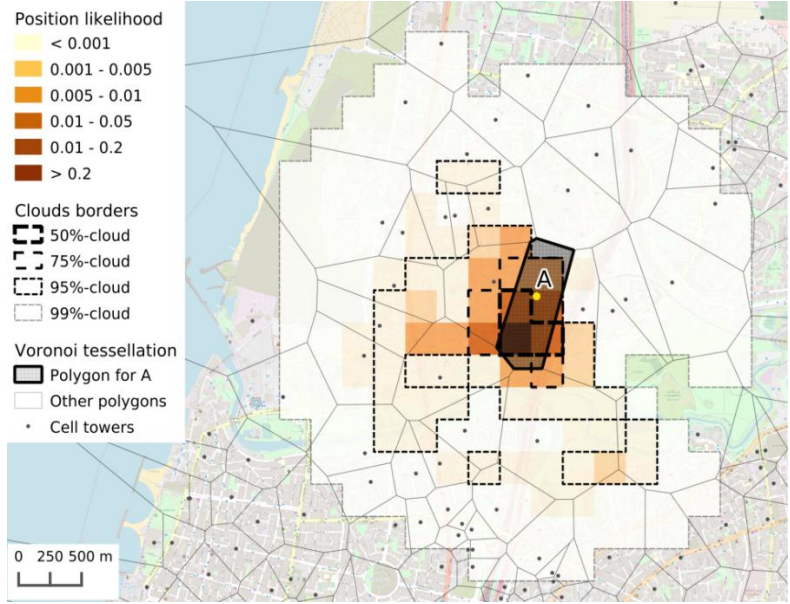

Figure 4: The p-clouds for cell tower M on a background of the MPN Voronoi coverage

We define *p-cloud for a tower M* as a minimal set of grid elements $\Delta_i$ with the highest value of the probabilities $P(\Delta_i|M)$ that comprise a total share greater than or equal to p.

Figure 4 presents the p-clouds for p = 0.5, 0.75, 0.95 and 0.99, for the tower M that serves the densely populated area of the Tel Aviv University campus on a background of the tower-based Voronoi partition. As can be seen, the tower's 75%- and larger clouds are much larger than tower's Voronoi polygon. At the same time, the top part of M's Voronoi polygon is hardly served by the antennas of this tower.

## 4. Comparison between the Bayesian and Voronoi-based positioning

In what follows we estimate the discrepancy between the Voronoi-based and Bayesian estimation of position for Partner's MPN. The MPN considered in our study consists of 2851 cell towers equipped with 22007 outdoor 3G antennas in total. Most of the antennas cover a 120° sector. To compare the deterministic Voronoi and the proposed Bayes locations, let us estimate some properties of the Partner's MPN Voronoi tessellation.

### 4.1. Statistics of the Voronoi tessellation

Figure 5 shows the distribution of the distance between a cell tower and its 4 closest neighboring towers in the MPN. As could be expected, the distribution of the nearest neighbor distance is essentially asymmetric.

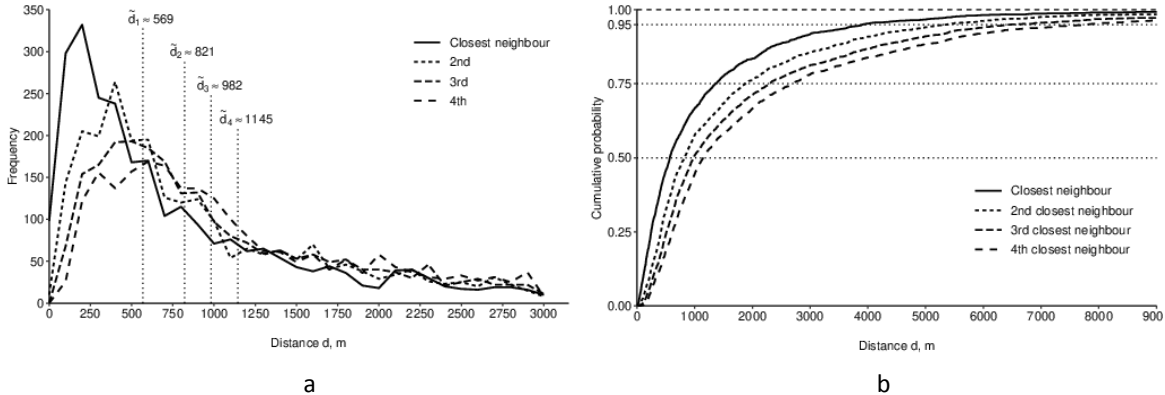

Figure 5: Distance to the nearest, $2^{nd}$, $3^{rd}$ and $4^{th}$ neighboring tower. (a) Distribution density for the distances below 3 km; (b) Cumulative distribution for the distances below 9 km; Vertical lines mark median distance

The size distribution of the Voronoi polygons has a very long tail (Figure 6). While the average polygon size is 9.89 km$^2$, only 20% of polygons are larger than this average, and half of the polygons have area less than 1.2 km$^2$.

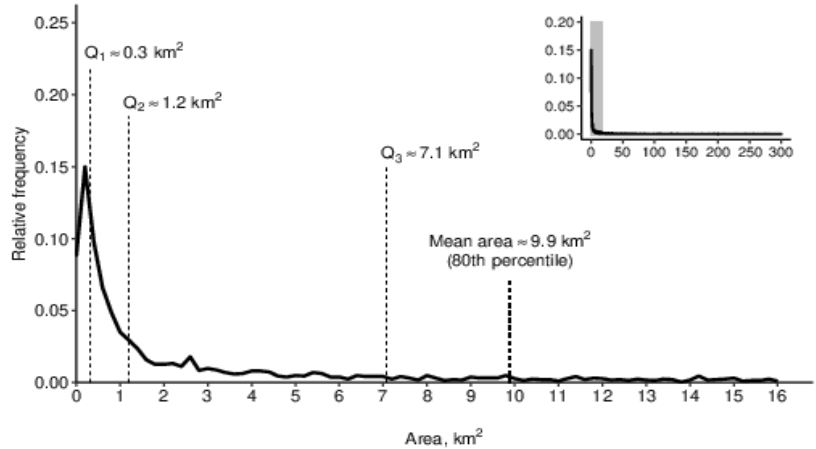

Figure 6: The PDF of the Voronoi polygons' size. $Q_1$, $Q_2$ and $Q_3$ denote the 25$^{th}$, 50$^{th}$, and 75$^{th}$ percentiles, respectively. The inset – full PDF, the size of the largest polygon is ~ 300 km$^2$

Despite high variation of the polygons' size, the MPN Voronoi partition remains close to a honeycomb: As can be seen in Figure 7, the number of neighbors for most of the polygons is between 5 and 7, similar to the Voronoi coverage constructed for the random Poisson point process. We thus conclude that an arbitrary point inside the tessellation is more likely to fall into a large Voronoi polygon than into a small one (Haenggi, 2013).

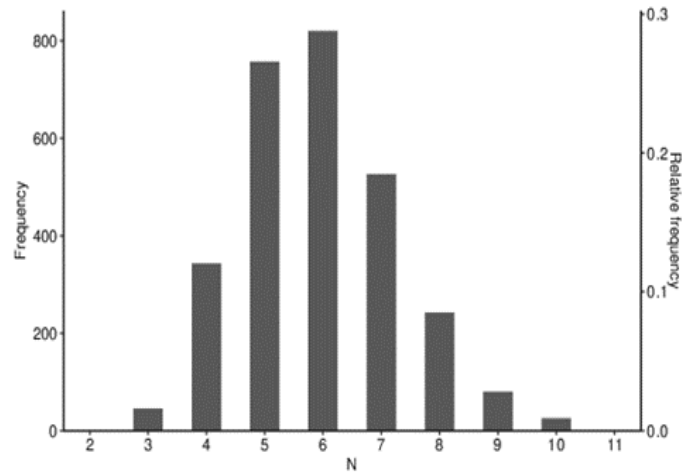

Figure 7. PDF of the number of neighboring polygons for Partner's MPN.

By design, such a tower pattern enables effective reuse of radio frequency bands, and increases the network's capacity and coverage area.

### 4.2. Statistics of the probabilistic clouds

Using the Bayesian model (1) – (3) we built the probabilistic cloud for every antenna – each element of a square grid is characterized by the probability to serve a device located in this cell for each antenna that covers it. The question is how large is the full probabilistic cloud and its p-clouds. In what follows, we consider p-clouds for p = 0.95, 0.75 and 0.50.

Cloud size and shape are dependent on the population pattern and topography, and we compare clouds' statistics for two regions of Israel: The densely populated lowland of the Tel Aviv Metropolitan Area and the averagely populated and hilly Haifa and Northern District (Haifa/North). According to the 2016 census, Tel Aviv Metropolitan Area (TAMA) has a population of 3.85 million people and its area is 1198 km$^2$, while the combined population of the Haifa/North region is 2.4 million and its area is 5494 km$^2$. These two regions also differ in shape. TAMA is located along the coastline and resembles a 50x20 km rectangle oriented south-north. The region of Haifa/North is close to square and has complicated landscape varying from the coast to the hilly highlands. As above, we consider antennas' TTBs with 10 or more connections per 250 × 250 m grid cell per month.

The variety of the overlap states between Voronoi and Bayesian coverages is very high. The average number of antennas serving the same grid cell according to the Bayesian clouds (Figure 8) is close to 32 for the 99%cloud, 12 for the 95%-cloud, 5 for the 75%-cloud and 2.5 for the 50%-cloud.

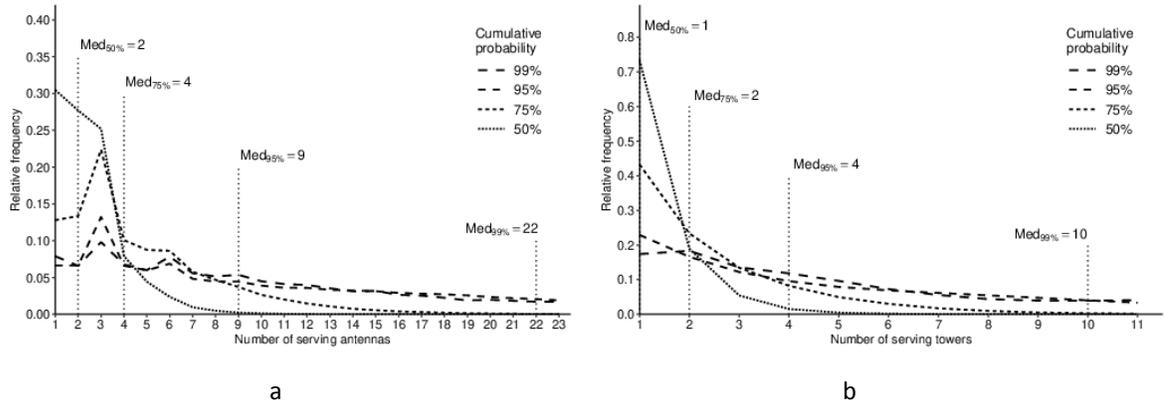

a  b

Figure 8: (a) Distribution of the number of antennas (up to 23) serving the same grid element for the p-clouds, p = 50, 75, 95, 99%. (b) The same distribution for the number of towers (up to 11)

The same estimates for the tower are twice lower (Table 2).

| Table 2. Number of antennas' and towers' p-clouds serving the same grid cell | | | | | | | | |
|---|---|---|---|---|---|---|---|---|
| | Antennas' p-clouds | | | | Towers' p-clouds | | | |
| p | 50% | 75% | 95% | 99% | 50% | 75% | 95% | 99% |
| Mean | 2.45 | 4.78 | 11.99 | 32.38 | 1.38 | 2.37 | 5.47 | 13.68 |
| SD | 1.47 | 3.47 | 9.88 | 32.14 | 0.74 | 1.79 | 4.64 | 13.13 |
| Median | 2 | 4 | 9 | 22 | 1 | 2 | 4 | 10 |
| IQR | 2 | 4 | 14 | 42 | 1 | 2 | 6 | 18 |
| Max | 18 | 33 | 85 | 343 | 9 | 18 | 38 | 125 |

Median size of the 95%-cloud is 2.5 km$^2$ (Figure 9a), 2 times larger than that of the tower's Voronoi polygon (1.2 km$^2$, Figure 6), while the average size of the 75%-cloud is close to the average Voronoi polygon size. The cumulative probability chart (Figure 9b) shows growth of cloud size from 0.5 km$^2$ for the 50%-clouds to 4.5 km$^2$ for 99%-clouds (Figure 9b).

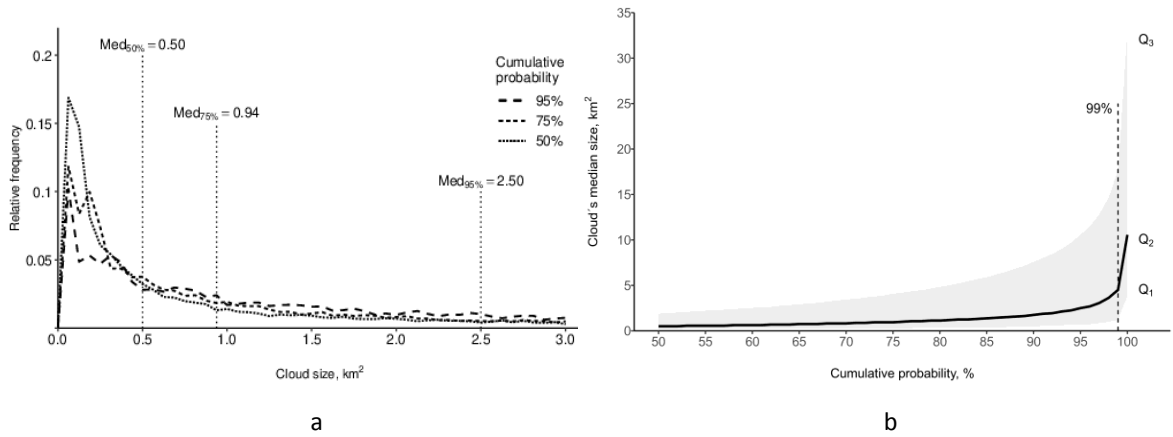

Figure 9: Distribution of the clouds size (a) and probabilities within the cloud (b). The shaded area in (b) depicts interquartile range with $Q_1, Q_2, Q_3$ for the 25th, 50th, and 75th percentiles.

### 4.3. The size of the towers' Bayesian clouds relative to the size of the Voronoi polygons

Figure 10 presents the ratio between the area of a tower's p-cloud and the area of the tower's Voronoi polygon. As can be seen, this ratio is similar in both regions and the size of the Voronoi polygon is, on average similar to the size of the p-cloud where p = 55-60%.

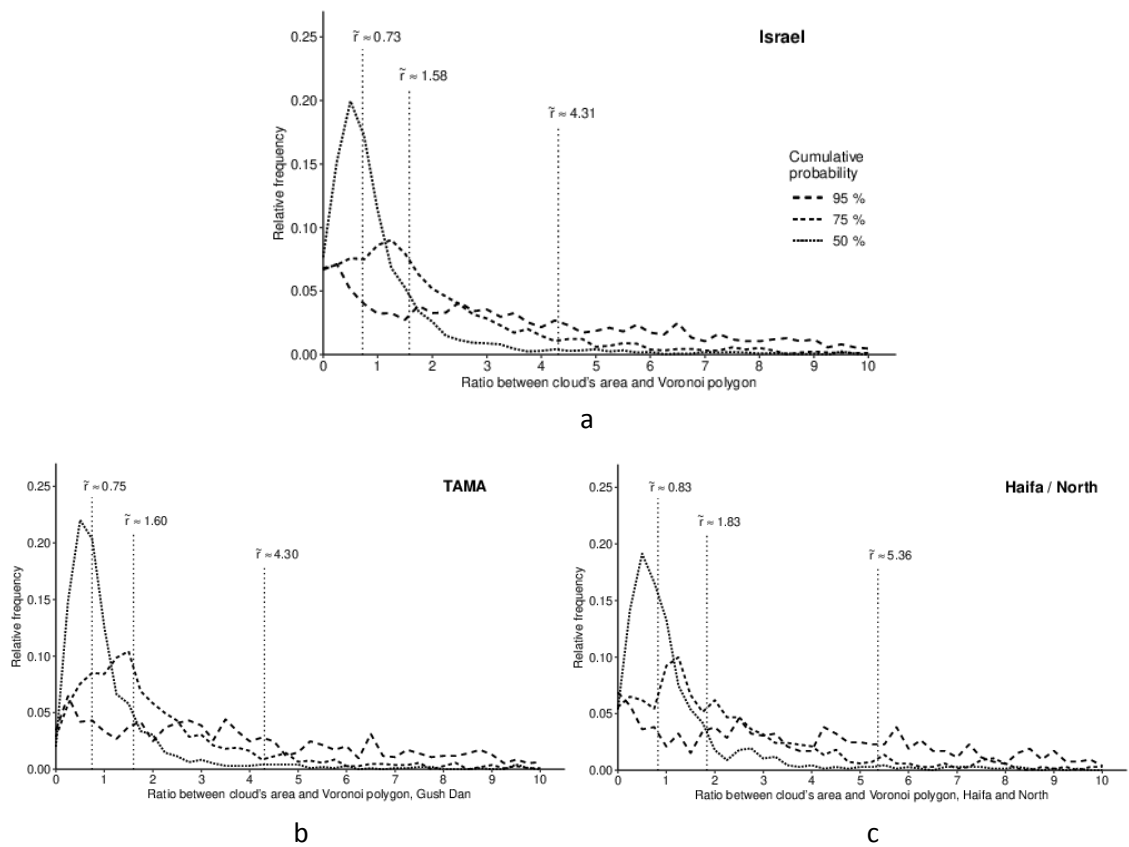

Figure 10: The ratio between the area of a tower's p-cloud and the area of the tower's Voronoi polygon for Israel (a) and two selected regions (b, c)

## 4.4. The overlap between the Voronoi polygons of clouds and towers

We represent the overlap by the probability that a device, registered by the antenna A which belongs to the tower M, is located in M's Voronoi polygon. To estimate this probability, we cut out from a 100%-cloud of M the part that is covered by the Voronoi polygon, and sum up the probabilities that characterize grid elements inside it.

As Figure 11 shows, the overlap covers the entire spectrum of options, from the polygons that do not intersect at all to the polygons that cover the entire cloud and even exceed it. The overlap differs in two regions and, for example, the polygons that contain the entire Bayesian cloud in Haifa/North comprise 3.6%, compared to 2% in TAMA. This can be explained by the larger distance between settlements in the North its less dense transport network: actively served areas may thus be smaller than the corresponding Voronoi polygons. Polygons that do not overlap with probabilistic clouds at all comprise 2.4% for Haifa/North region and 2.5% for TAMA region. Except for the extremes, Voronoi polygon is most likely to cover between 25 and 50% of the Bayesian cloud.

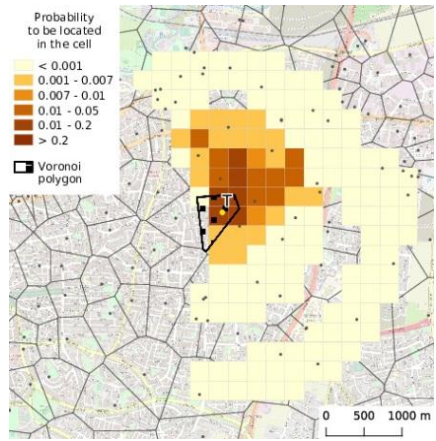

a

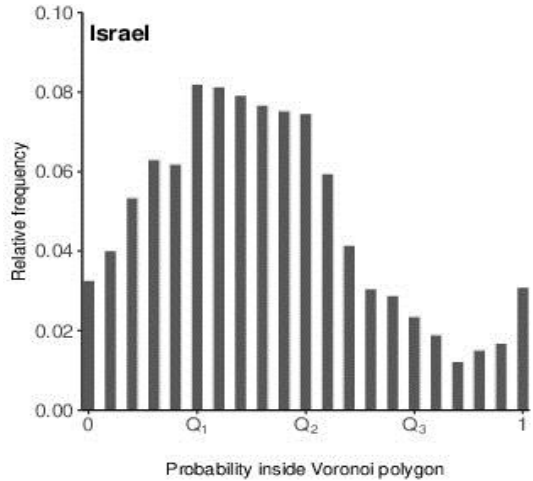

b

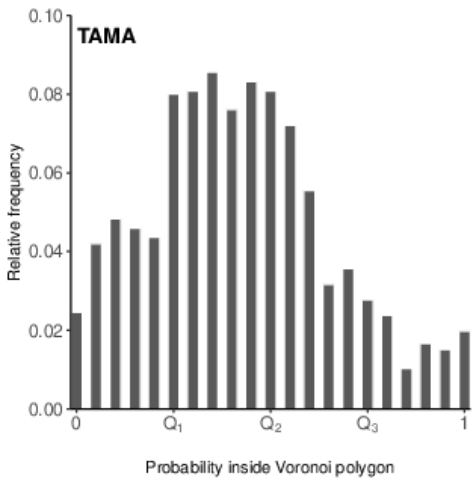

c

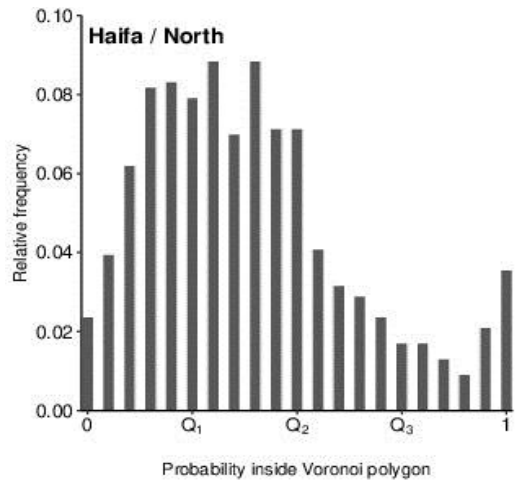

d

Figure 11: The cumulative probability within the part of the tower's Bayesian cloud that falls inside the tower's Voronoi polygon. (a) An example of the Voronoi polygon that accumulates 35% of the overall tower's cloud probability (b) Entire Israel; (c) TAMA (d) Haifa/North

## 5. Consequences of probabilistic positioning for the location privacy

A shift from the traditional Voronoi-based technique to the probabilistic Bayesian location inference has significant consequences for the assessment of privacy of mobile device users. Namely, the overlap between the service areas of antennas results in an essentially lower certainty of locating a device than is stated according to the Voronoi-based view of location.

Let us consider a typical scenario of the attack on mobility privacy: An adversary intends to reveal the mobility history of target person *p* who uses the device D, using both surveillance and access to the anonymized CDR dataset $\mathcal{A}$. Anonymization in this particular case means decoding or removing from $\mathcal{A}$ any personal information that can be known by the adversary from other sources. We assume that the adversary 1) can identify the location of the target by direct observation, with finite accuracy ε and 2) possesses a full knowledge of the MPN towers and antennas and their mapping into the records of $\mathcal{A}$.

The exact format of the dataset $\mathcal{A}$ depends on the data processing workflow designed to fulfil the specific business goals of the cellular company. It is usually assumed that $\mathcal{A}$ consists of CDRs that contain device ID, time of connection establishment, time of connection termination, cell tower ID, and antenna ID. Typically, more than one CDR is recorded during the voice call. If the call is managed by several antennas, a new CDR is created for every segment of the call (communication session).

Traditionally the location of D is based on the Voronoi coverage that is constructed based on the cell towers' locations. Namely, if D's CDR is recorded during the communication session, then it contains the ID of the antenna and cell tower C, and D can be located within the Voronoi polygon $V_C$ of C (Figure 12, (De Montjoye et al. 2013)). Often, the location of C itself is considered as a proxy for D's location.

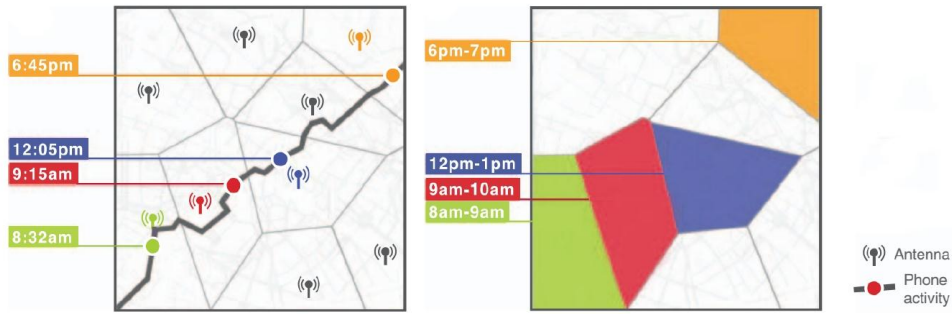

Figure 12. (a) Example of the individual trace, the dots represent the locations during communication sessions. (b) Voronoi-based view of the same trace, (De Montjoye et al. 2013).

To recognize D's records in ⊿, the adversary starts with locating the target by direct observation. Knowing target's position at a moment $T_0$, the adversary, based on knowledge of the MPN, determines the corresponding Voronoi polygon $V_C$ and then the cell tower C. Further, the adversary queries ⊿ for all CDRs of the communications performed through C within the interval of time $[T_0 - \Delta t, T_0 + \Delta t]$, where the uncertainty in the time condition is defined by the known inaccuracy of a time measurement and delay in connections processing.

Being sure of Voronoi-based positioning, the adversary believes that D has to be among the query's result. If we denote this result as $Q_0$ and the set of candidate devices' IDs by $C_0$, the uniqueness of the target device p can be estimated as:

$$U_0(p) = \frac{1}{|C_0|}.$$

If $U_0(p) = 1$, the attack is successful. Otherwise the adversary continues the surveillance and at the moment $T_1$ determines Voronoi cell $V_1$ and queries ⊿ again. He believes that target's phone ID has to be in $C_0 \cap C_1$ and target's uniqueness is

$$U_1(p) = \frac{1}{|C_0 \cap C_1|}.$$

The adversary can proceed, obtaining candidate sets $C_2, C_3, \ldots$ and target's uniqueness estimates

$$U_k(p) = \frac{1}{|C_0 \cap C_1 \cap \ldots \cap C_k|}.$$

Since $|C_0 \cap C_1 \cap \ldots \cap C_k| \geq |C_0 \cap C_1 \cap \ldots \cap C_{k+1}|$, $U_0 \leq U_1 \leq U_2 \leq \cdots$. Empirical tests of $U_k(p)$ values for real devices demonstrate that the sequence of 3 - 5 records with different towers, even without the time tags, is sufficient to identify almost 99% of the devices (De Montjoye et al. 2013).

The above line of argument is based on the assumption that device D, whose communication session was performed via cell tower T, is located within the $V_T$. As we demonstrate, the devices that were served by T are located in an area that is several times larger than $V_T$, and this area is simultaneously served by many other antennas located on many towers.

To sum up, Bayesian estimates of a device's location proposed in this paper result in a significant increase in the possible area of device position, and the probability that the device is located beyond the tower's Voronoi polygon is high. We thus conclude that the traditional Voronoi-based approach to the location privacy of mobile devices is essentially overcautious, and we will further investigate methods for locating the mobile devices in a consequent paper.

## Acknowledgements

The research was funded by the Blavatnik Foundation grant and performed in the framework of the privacy preserving agreement between TAU and Partner communication LTD. A.O. is partially supported by the Israeli Ministry of Aliyah and Integration.

The authors are grateful to Michal Ferenz for her assistance in the research and to Partner Communications LTD and Partner's Data Science group staff – team leader Riki Zuriel, data scientists Achinoam Soroker and Alex Kosov, data engineer Itzik Vakshi, and David Levy, radio manager.

Many thanks to Carole Shoval for the language editing.

## Author contributions statement

A.O. proposed the idea of the Bayesian positioning and performed the computations. I.B. headed the research and participated in data analysis. A.O. and I.B. wrote the paper. I.O. prepared the literature review. B.A. established the CDR database for analysis at Partner Communications LTD. All authors reviewed the manuscript.